\documentclass[a4paper]{jpconf}
\usepackage{graphicx}
\begin{document}
\title{A framework to monitor activities of satellite data processing in real-time}

\author{M D Nguyen$^1$, A P Kryukov$^1$}
\address{$^1$ Skobeltsyn Institute of Nuclear Physics, Lomonosov Moscow State University, Moscow, Russia}
\ead{nguyendmitri@gmail.com}

\begin{abstract}
Space Monitoring Data Center (SMDC) of SINP MSU is one of the several centers in the world that collects data on the radiational conditions in near-Earth orbit from various Russian (Lomonosov, Electro-L1, Electro-L2, Meteor-M1, Meteor-M2, etc.) and foreign (GOES 13, GOES 15, ACE, SDO, etc.) satellites. The primary purposes of SMDC are: aggregating heterogeneous data from different sources;  providing a unified interface for data retrieval, visualization, analysis, as well as development and testing new space weather models; and controlling the correctness and completeness of data. Space weather models rely on data provided by SMDC to produce forecasts. Therefore, monitoring the whole data processing cycle is crucial for further success in the modeling of physical processes in near-Earth orbit based on the collected data. To solve the problem described above, we have developed a framework called Live Monitor at SMDC. Live Monitor allows watching all stages and program components involved in each data processing cycle. All activities of each stage are logged by Live Monitor and shown in real-time on a web interface. When an error occurs, a notification message will be sent to satellite operators via email and the Telegram messenger service so that they could take measures in time. The Live Monitor's API can be used to create a customized monitoring service with minimum coding.
\end{abstract}

\section{Introduction}
One of the most critical tasks of a space monitoring data centre is providing correct data collected from satellites which are suitable to be used in research. The data collected from satellites are usually called the raw data. These raw data need to be processed and converted into a proper format (CSV, TXT, JSON, CDF, HDF, etc.) so that they can be read by analytical programs. An automatic data processing, storage and distribution system called SDDS has been created in 2016 at Skobeltsyn Institute of Nuclear Physics for this purpose. SDDS automates the whole cycle of satellite data processing which consists of the following steps: 1) connecting to data sources of each satellite; 2) checking for new data; 3) downloading them to a temporary storage;
4) decoding data if they were encoded; 5) extracting raw instrumental data from decoded data; 6) producing scientific data suitable for further analysis from the raw ones; 7) inserting scientific data into a unified database; 8) moving both raw and scientific data to long-term data storage with compression on demand. In each step, a set of programs are involved. Processed data are considered to be correct if every program involved in processing completed with no error. So to be sure of the data correctness, we need a mechanism that can monitor all activities of each program during a processing cycle and produce a report at the end. If an error occurs during a processing cycle, the mechanism must detect the component that caused the error and inform satellite operators and developers immediately so that they could take measure in time.

To solve the task we have developed a monitoring subsystem called Live Monitor. Later we decided to transform Live Monitor into an independent framework with RESTful API so that other satellite developers could build their own monitoring system based on Live Monitor.

This paper is organized as follows. In the second section, we consider several existing solutions to the problem compared to our. In section 3, we give a more detailed view of the overall architecture of the Live Monitor framework and show how it works. In section 4, we demonstrate an use case where the framework was used as the monitoring system in processing data of the Meteor-M2 satellite. In conclusion, we give a short resume of our completed work and describe our vision of the future perspective.

\section{Related works}
\label{related_works}
Our framework has been developed for use mostly in Linux operating systems. In practice, there are two ways to monitor activities of a running program: the passive way and the active way. The idea of the passive way is as follows. There is a master server running permanently on a machine and waiting for incoming requests. This server also serves as a web server for displaying statistics and a notification server to broadcast messages to subscribed users. On the machine where the target program is running there is another slave server running. The slave server triggers a checking script to get the current condition of the target program and send it to the master server on a regular basis. The idea of the active way, so-called lightweight event-driven or push notification, is that whenever an event arises the target program sends a short message (or a signal) to a master server, and the master server, in turn, broadcasts the message to all subscribed users.

Popular IT infrastructure monitoring solutions, such as Zabbix \cite{zabbix}, Nagios \cite{nagios}, MMonit \cite{mmonit}, and collectd \cite{collectd}, use the passive way to gather metrics of target programs. In our case, to monitor a program involved in data processing using the passive way we need to write a wrapper which runs a number of tests and returns certain metrics as a result. The client monitoring server then triggers the wrapper in a regular basis, for example, a 5-minute interval, to check the target program. This approach is not suitable for us because of the delay time between checks. Writing a wrapper for each target program would lead to a big amount of source code to be maintained. Furthermore, additional checks on a regular basis will produce overhead and affect the overall performance of the operating system.

On the other hand, push notification would inform us about the current condition of the target program immediately when an event arises. Currently, there are many available push notification solutions with API ready in the market such as Amazon Simple Notification Service \cite{amazon}, Urban Airship \cite{urban}, Appsfire \cite{appsfire}, One Signal \cite{onesignal} and so on. The primary problem is that most of these solutions are commercial. Free solutions often imply different constraints on how it can be used: a limited number of message queues or a limited amount of messages generated by one user. One Signal, for example, requires that one must have a website with a public domain and messages must be generated only by the application bound to the website. Appsfire provides SDK only for mobile platforms. Thus, we decided to create our own solution - a flexible framework with an open API and no constraint.

\section{Live Monitor's architecture}
\label{live_monitor_architecture}
Live Monitor framework consists of the following components: a backend logging library, RabbitMQ message broker, a RESTful API backend, a frontend UI library. An illustration of the architecture is shown below in figure \ref{figure_architecture}. In the figure, the target programs to be monitored are components of the SDDS system such as satellite controllers, instrument data decoders, DB loader, and others.

\begin{figure}[ht]
\centering
\includegraphics[height=20pc, clip]{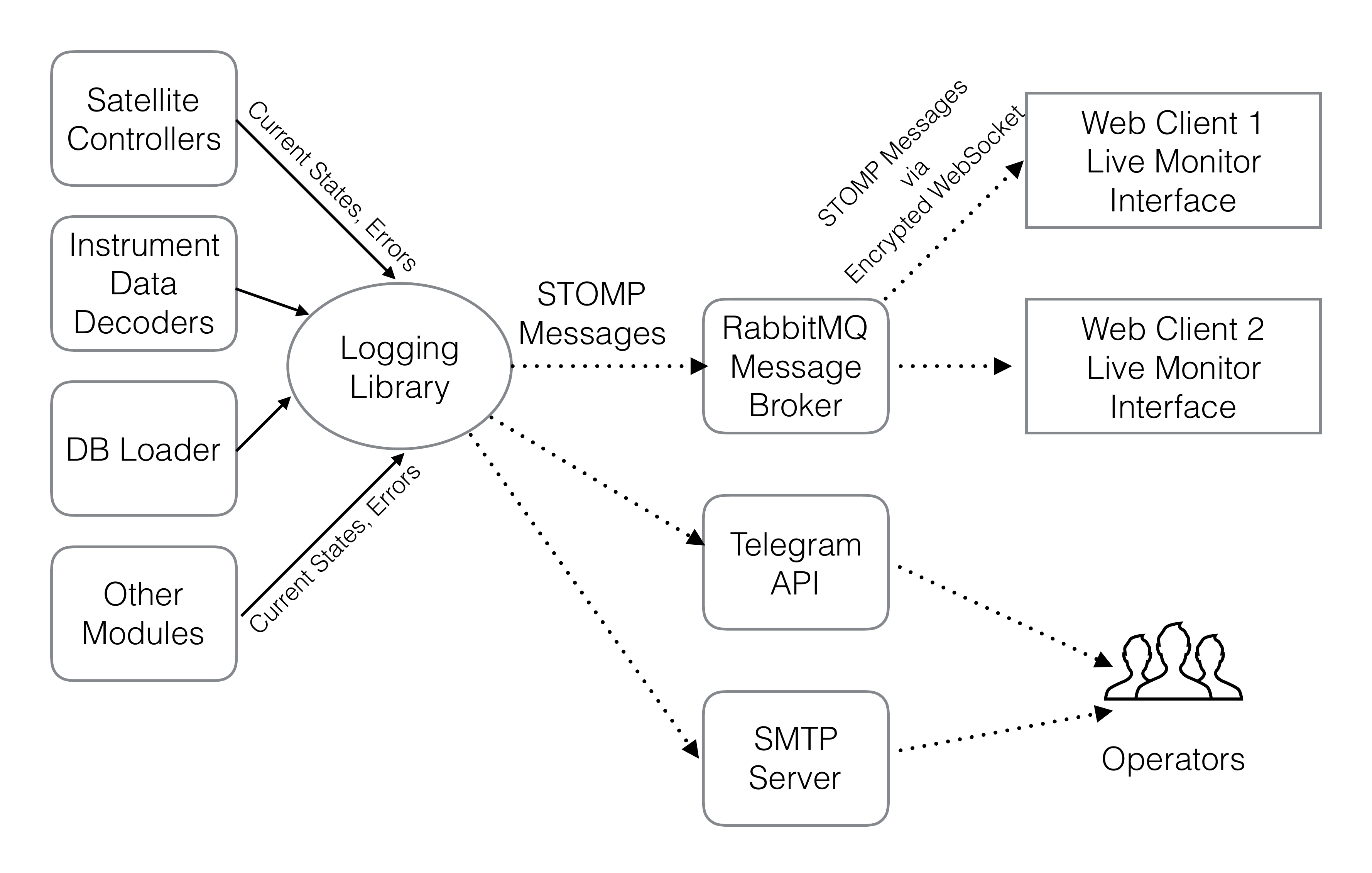}
\caption{Live Monitor's architecture}
\label{figure_architecture}
\end{figure}

The backend logging library is based on the standard Python Logging module. The logging library supports four levels of logging messages: debug, info, warning, and error. Error messages are logged when an error occurs during the data processing cycle which could lead to incorrectly processed data or cause a component failure. Error messages are also sent directly to the operators via the Telegram messenger service and/or email. Warning messages are logged for minor errors that do not affect the data correctness and normal functioning. Info messages are just normal text descriptions of events during the data processing cycle. Debug messages include diagnostic information that is helpful in failure investigation. Besides the standard behaviour which is writing short text messages in different log levels to a log file, the logging library sends these text messages to the RabbitMQ message broker \cite{rabbitmq} via a TCP socket. Messages are formatted using the STOMP protocol \cite{stomp}. RabbitMQ, in turn, broadcasts received text messages to all subscribed frontend clients that use the frontend UI library to display messages on web pages. Notification features can be turned on and off by editing proper configuration files or using the RESTful API. Currently, the following operations are supported by the RESTful API:
\begin{itemize}
\item create/delete a customised monitoring service;
\item change the logging level;
\item switch the whole monitoring service on/off;
\item switch a component of the monitoring service on/off;
\item switch Telegram message delivery for a service on/off;
\item switch e-mail delivery on/off.
\end{itemize}

\begin{figure}[ht]
\centering
\includegraphics[height=16pc, clip]{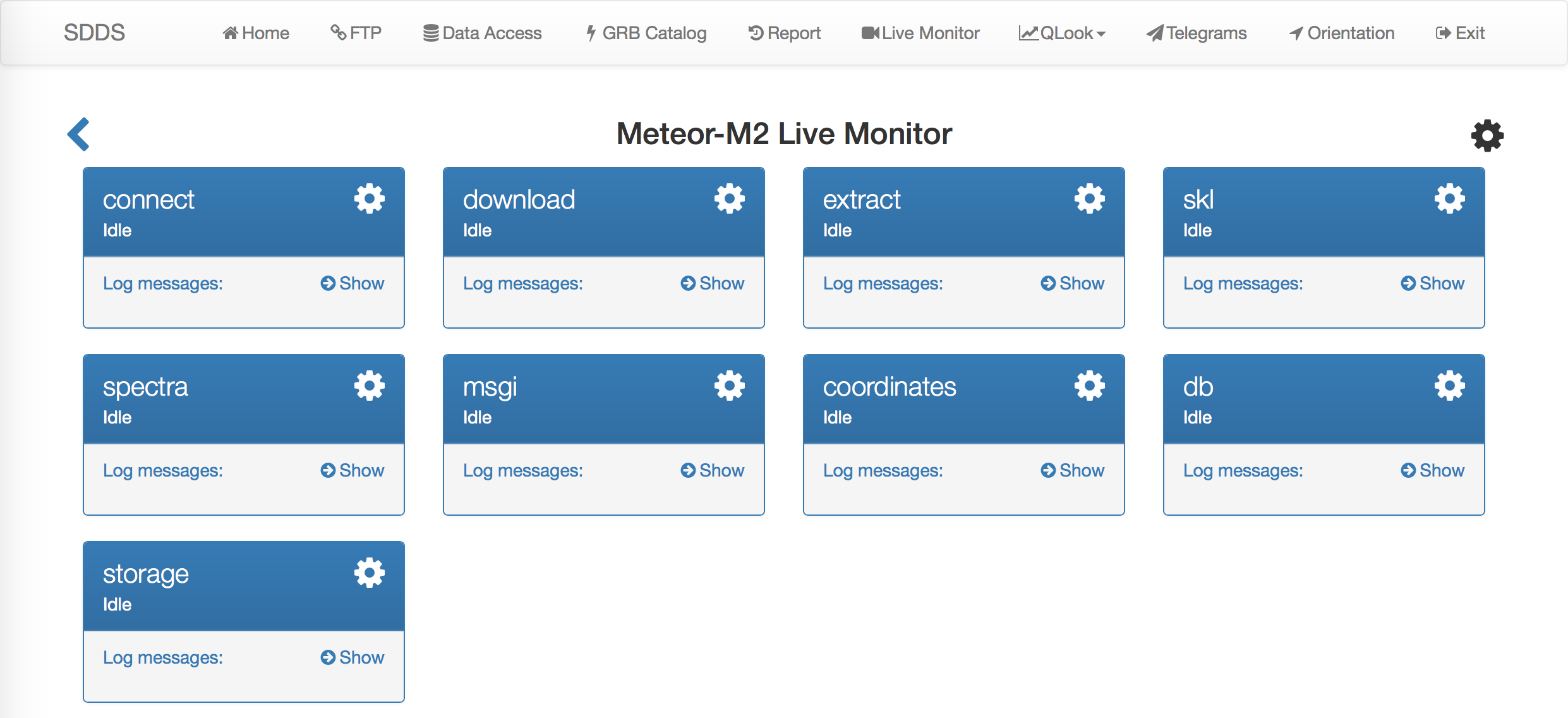}
\caption{The data processing cycle of Meteor-M2}
\label{figure_web_ui}
\end{figure}

\begin{figure}[ht]
\centering
\includegraphics[height=16pc, clip]{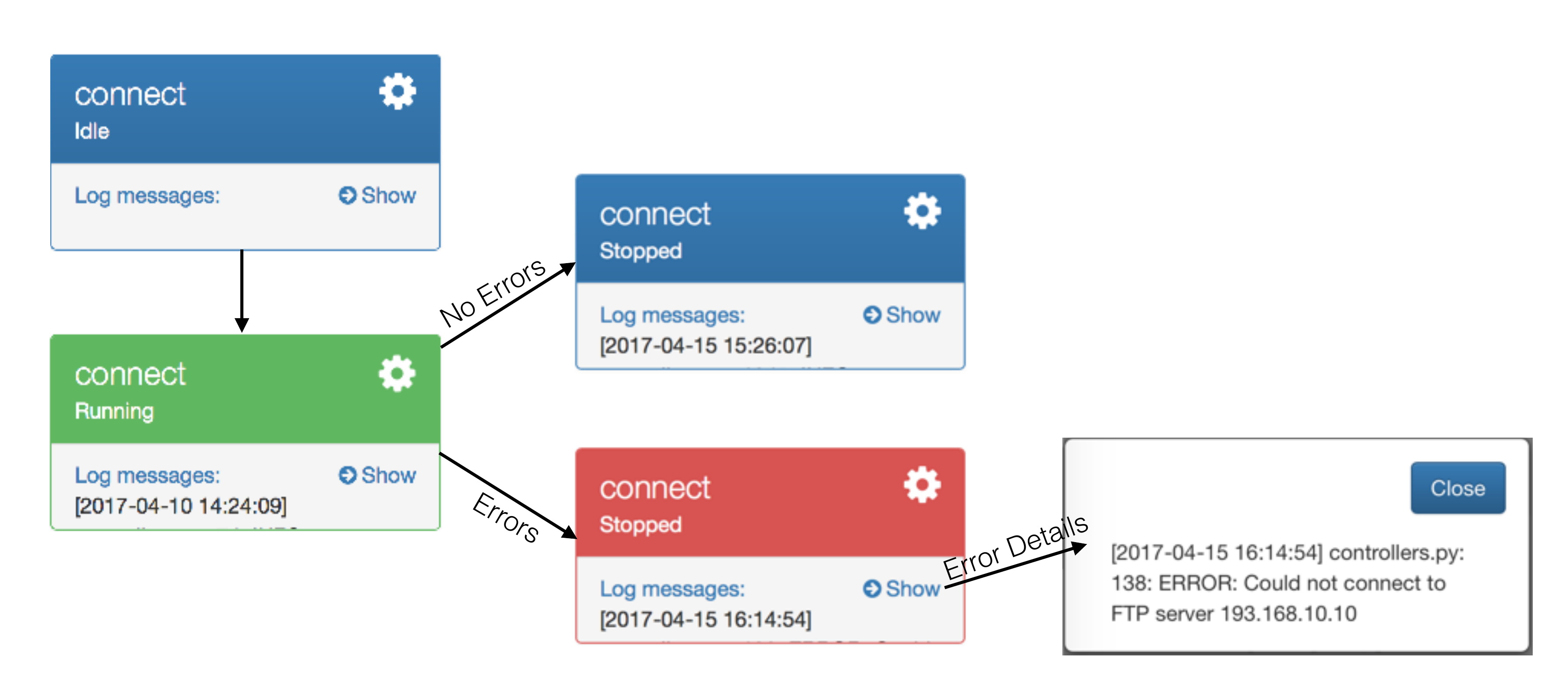}
\caption{Different states of the “connect” step of Meteor-M2}
\label{figure_statuses}
\end{figure}

The main goal of the frontend library is to control how each step of a data processing cycle should be shown on the web interface. The frontend library sends a GET request to the backend to retrieve necessary information of what should be shown. The answer from the backend is a JSON object that consists of a number of steps. After that, the frontend library establishes a WebSocket \cite{websocket} connection with the RabbitMQ message broker. When a text message of a step arrives, the frontend library parses its content and changes the visual appearance of the step. In figure \ref{figure_web_ui} and figure \ref{figure_statuses} all steps in the processing cycle of the Meteor-M2 satellite and possible states of the \verb“connect” step are illustrated.

\section{Monitoring Meteor-M2 using Live Monitor}
\label{meteor_m2}

In this section, we demonstrate how we used Live Monitor to track activities in data processing of the Meteor-M2 satellite. To register the monitoring service, we sent a POST request to the backend. We described the monitoring service as an object using the JSON format and passed the object as the data parameter of the request. The backend generated a JSON configuration file according to input parameters and created a new message queue.

\begin{verbatim}
  # POST request parameters
  {
    "satellite": "meteor_m2",
    "instruments": ["msgi", "skl"],
    "norad_id": 40069,
    "sources": {...}
  }

  # a part of the generated configuration file
  ...
  "notification": {
    "email_enabled": true,
    "email": "***",
    "smtp_server": "***",
    "smtp_protocol": "tls",
    "smtp_port": 587,
    "smtp_user": "***",
    "smtp_pass": "***",
    "push_enabled": true,
    "push_host": "localhost",
    "push_port": 61613,
    "push_user": "***",
    "push_pass": "***"
    "push_queue": "/topic/meteor_m2",
    "telegram_enabled": true
  },
  ...
\end{verbatim}

In each target program, firstly, we created a logger object using the get\_logger function provided by the backend library. Then whenever we need to inform about an event, we called the logger object with either debug, info, warning, or error functions. It is possible to add a customised text as a parameter to the function. The logging library will format the text and send it further.

\begin{verbatim}
  # importing the logging library
  import log
  ...
  # creating the logger
  self.logger = log.get_logger( module_name = module_name,
                                stomp_cfg = cfg['notification'],
                                telegram = True,
                                stdout = True,
                                log_file = log_file,
                                log_level = log_level )
  ...
  # using the logger to log event
  self.logger.error('Error running decoder, return code = %r' % ret_code)
\end{verbatim}
On the frontend side, we created a web page and included the Javascript library provided by Live Monitor with authentication credentials taken from the generated JSON configuration file as well as WebSocket and STOMP libraries. When we open the web page, the Live Monitor library makes a request to get the configuration that describes the appearance of notification messages, initialises a WebSocket connection to the RabbitMQ STOMP server, and subscribes to the message queue. Each time after receiving a new event from the message queue the Live Monitor library will trigger a proper function to handle the event and to show it on the web page.

\section{Conclusion}
\label{conclusion}
The Live Monitoring framework is now actively used in developing new components of the satellite data processing system (SDDS) at SINP MSU. For 18 months of operation, Live Monitor has been helping us identify and localise the scope of any occurred problem in data processing in time and hence prevent or fix the problem quickly. We conducted a stress test which showed that Live Monitor was able to deliver up to 10000 messages per second on the following hardware: E5-2650 2.60GHz / 8GB memory / 1Gb Ethernet. In future, we plan to support more operations to control the behaviour of monitoring services such as changing components of a monitoring service dynamically and controlling the service via Telegram Bot.

\section*{Acknowledgments}
We would like to thank Dr Vladimir Kalegaev for helpful discussions on satellite data processing and clear problem statements. This project was supported by RSF grant \#16-17-00098.

\end{document}